# Dynamic Functional Connectivity

Christine Ahrends[1] and Diego Vidaurre[1,2]

[1] Center of Functionally Integrative Neuroscience, Department of Clinical Medicine, Aarhus University, Denmark

[2] Department of Psychiatry, University of Oxford, United Kingdom

## Abstract

Most generally, dynamic functional connectivity (FC) refers to the non-instantaneous couplings across timeseries from a set of brain areas, here as measured by fMRI. This is in contrast to static FC, which is defined as purely instantaneous relations. In this chapter, we provide a hands-on description of a non-exhaustive selection of different methods used to estimate dynamic FC (such as sliding windows, clustering approaches, Hidden Markov Models, and multivariate autoregressive models), and we explain, using practical examples, how data should be prepared for dynamic FC analyses and how models of dynamic FC can be evaluated. We also discuss current developments in the dynamic FC research field, including challenges of reliability and reproducibility, and perspectives of using dynamic FC for prediction.

## 1  Background & motivation

The most traditional use of fMRI in research has been to detect changes in the signal according to some task design, for example in response to a visual stimulus (Filippi, 2016; Frahm et al., 1992; Friston et al., 2006; Kwong et al., 1992; Ogawa et al., 1992). This is a first-order statistic of the signal —i.e. the average value of the signal for a given region. Since the scale of the BOLD signal is not necessarily interpretable, it makes sense only relative to a baseline. For example, is the signal in a given region higher on average after a stimulus presentation than before?

In the last decade, however, there has been an explosion of studies aiming to characterise individual differences in fMRI activity, not only in task but also in rest —that is, in the absence of an explicit, constrained task (Lee et al., 2013; Lurie et al., 2019; Smith, Vidaurre, et al., 2013). In rest, first order statistics are less useful because there is no baseline to compare to, and there are no observable conditions to build a contrast. This is one of the reasons why we look at second-order statistics, such as functional connectivity (FC), which can be interpreted above and beyond the (arbitrary) scale of the signal (Power et al., 2011). Another reason is that we expect FC to capture aspects of neural function and structure that are not directly linked to, or visible in, first-order statistics. One example is neural communication, which however might be reflected in FC in ways that are not necessarily direct or linear. FC can also reveal aspects of topographical organisation at large brain scales that cannot be retrieved from first-order statistics (Margulies et al., 2016).

FC is defined as a statistical relationship between two brain areas. The most common measures of FC are linear, namely Pearson's correlation or unnormalised covariance, or the inverse of these, which serves to partialise the information; that is, in the inverse of a covariance matrix (also called a precision matrix), the coefficient relating area $i$ and area $j$ defines their (linear) relation after removing the influence of all the other areas (Joseph F. Hair, 2009; Joseph F Hair, 2009). There are other measures of FC that aim at capturing nonlinear information, most remarkably based in information-theory, such as mutual information. Although these measures are more general and can potentially capture other aspects of connectivity that escape a covariance or correlation matrix, they either require more data to obtain a clean estimation or impose their own assumptions. In this chapter, we will focus our discussion on linear FC, since it is by far the most common choice, not just in general but particularly when the interest is in characterising the dynamic aspects of FC.

Another useful distinction is between FC and effective connectivity (EC) (Friston, 2011). As opposed to FC, which is a mere statistical description of the data, EC is defined as the direct, causal influence that an area has over another in the context of the network to which they belong, and under specific models of biophysical constraints. Whereas EC can capture aspects of the data that FC cannot, the estimation of EC is also more dependent on specific assumptions and the inference of the parameters is more difficult. This chapter only focuses on FC.

But what is dynamic FC precisely? Most broadly, we can define it as any second-order, cross-region information that is not captured by the so-called static FC. Static FC is a usual term to refer to the correlation between voxels or regions as computed within the entire duration of the scanning session. Note that, since the standard practice is to standardise the signal per scanning session to have a mean zero and standard deviation one, the covariance and correlation matrices are mathematically equal when we speak about static FC. Critically, static FC, as per this definition, only captures instantaneous relationships; that is, within a scanning session, if we permute the time points within a session such that the permutation is the same for all voxels or areas, the static FC estimate would not change. But there is other information that would be lost by permuting; this information exists above and beyond the static FC, and that is what we refer to as dynamic FC here.

One of the problems in the study of dynamic FC is that the literature uses the same term for different types of measures, which has considerably muddled the discussion about the topic. By basing our discussion on a broad definition of dynamic FC, we intend to clarify the different aspects of dynamic FC that the literature has covered. Under the umbrella of this definition, there are two different aspects of dynamic FC: time-varying instantaneous FC, and FC in the context of a linear dynamical system that models non-instantaneous aspects of FC. For now, we will discuss them conceptually. Later, when we cover the different approaches to model dynamic FC, we will specify which type of information each approach is aiming to model.

Time-varying instantaneous FC, commonly referred to as just time-varying FC, reflects within-session modulations of the covariance matrix across regions, which captures between-area linear couplings. We note that, after standardising the signal, the mean is zero and the variance is one across the entire session, but this is not necessarily the case for shorter periods of the signal, where the signal, for example, could transiently have a higher variance. This means that covariance and correlation are not exactly equivalent anymore, opening different possibilities in how we model and interpret the data. These practical aspects and what they mean conceptually will also be discussed below.

Apart from instantaneous FC, the signal has other temporal information at different time scales, including the effect imposed by the hemodynamic response function as well as other factors of neural and non-neural origins. These non-instantaneous couplings between areas, which are also dynamic FC, can be modelled as a linear dynamical system, like the autoregressive model. In short, an autoregressive model models the multivariate signal at time point $t$ (namely $X_t$) as a linear function of previous time points plus (Gaussian) noise. This linear function is embodied by a set of autoregressive coefficients *A*, which, however linear, contain rich information about the data. We will cover autoregressive models in some detail in the *Practical Approaches to dynamic FC* section. For now, it suffices to say that A, with a single set of parameters, can capture non-instantaneous phenomena such as travelling waves, chaotic behaviour, and oscillations.

Different approaches of analysis can capture one or the other aspect, or both. But importantly, they are not independent from each other. For example, if the signal contains both types of information, but we use an analysis approach that is focussed on only one, information of the other aspect will necessarily leak into the estimation. For this reason, they are not trivial to separate, and arguments about which one is a better-grounded description of the data are not easy to make. More practically, different approaches to dynamic FC allow us to address different research questions in complementary ways; see Calhoun et al.

(2014); Hutchison et al. (2013); Lurie et al. (2019); Preti et al. (2017) for some examples of applications.

This chapter is mainly devoted to practical aspects of dynamic FC analysis. This includes: data preparation prior to the analysis, a non-exhaustive description of existing approaches to characterise dynamic FC (with an emphasis on implementation), and some discussion on how to validate these models after they have been estimated, followed by conclusions and perspectives. Where applicable, we will demonstrate examples with Matlab code or refer to existing software packages that can be used for the implementation. We chose Matlab because it may still be the most common language in neuroimaging, but a translation of the code examples to other languages like Python or Julia is straightforward. All code examples can also be found at https://github.com/ahrends/DynamicFC_examples.

# 2  Preparing data for dynamic FC analysis

When preparing fMRI data for dynamic FC analyses, a good starting point is to follow preprocessing guidelines for resting state fMRI, such as the Human Connectome Project's (HCP) resting state preprocessing pipeline (Glasser et al., 2013; Smith, Beckmann, et al., 2013). Resting state preprocessing guidelines may be more suitable than task-specific preprocessing recommendations since traditional analyses of task data typically take advantage of averaging over trials. There are a few considerations specific to dynamic FC approaches, which we will outline here. For further reference, the issue of preprocessing for dynamic FC analyses has also been tested and discussed in Lydon-Staley et al. (2019) and Vergara et al. (2017).

## 2.1  Temporal preprocessing

Most importantly, dynamic approaches are more heavily affected by temporal noise than time-averaged types of analyses. With temporal noise, we here refer to any type of artefact that varies over time, e.g., head motion, other physiological artefacts such as cardiac or respiratory artefacts, etc. While these artefacts may almost disappear when averaging over timepoints, such as in time-averaged or trial-averaged approaches, they can drastically influence the estimation of dynamic FC (Nalci et al., 2019). For instance, if the dominant temporal fluctuations in an fMRI timeseries are due to head motion, a dynamic FC model may use its explanatory power to describe these movement-related variations rather than the more subtle signal fluctuations stemming from neural activity. In a state-based model, this may result in one or more states being actually motion states and not "brain" states, while in a continuously-varying FC estimation, each FC estimate may to some extent be biased by motion.

On the other hand, dynamic FC analyses also suffer from too aggressive clean-up in the time domain, as some of the meaningful temporal variability can be removed along with the temporal artefacts. This may be the case, for instance, by applying preprocessing approaches that average over or censor time points, such as motion scrubbing (Power et al., 2012), or when regressing out temporal noise components using a full variance clean-up approach. Also global signal regression can affect temporal variability, both positively and negatively, and so its use in preprocessing data for dynamic FC analysis is controversial (Murphy & Fox, 2017).

The goal in terms of temporal variability when preprocessing fMRI data for dynamic FC analyses should therefore be to remove temporal artefacts while retaining non-artefactual temporal variability as much as possible. This may be achieved by non-aggressive temporal preprocessing strategies such as independent component analysis (ICA) in combination with unique variance clean-up of noise-related components (Griffanti et al., 2014). An additional consideration as regards temporal variability is temporal filtering. While a relatively lenient high-pass filter can be useful in removing ultra-slow fluctuations, such as scanner drifts, a very narrow filter will restrict the timescale on which dynamic changes in FC can be detected by a model. In general, since temporal noise and the meaningful aspect of the signal are not perfectly separable, achieving the right balance can be a difficult task. A sensible approach may be to start with a relatively lenient temporal clean-up and to test post-hoc whether the dynamic FC model was affected by known temporal artefacts, such as head motion, and decide whether more aggressive temporal clean-up is necessary (Ciric et al., 2017; Parkes et al., 2018).

An additional factor when doing dynamic FC analyses, and in particular time-varying FC analyses, is the variability of time-averaged FC between subjects; i.e. how different subjects are in terms of their average between-region correlations. Whether this variability between subjects is artefactual (e.g., due to poor registration) or non-artefactual (i.e. meaningful individual differences), it may mask the more subtle temporal variations in FC, making dynamic FC difficult to detect in a group-level timeseries (Lehmann et al., 2017). While non-artefactual between-subject variability may be of interest in the analyses, preprocessing should strive to minimise artefactual differences between subjects when planning a group-level dynamic FC analysis. In this case, it is recommended to test how representative FC patterns are of the group of subjects. A poor balance of low temporal variability and high between-subject variability can lead a dynamic FC model to converge to a static solution, which only describes differences between subjects (Ahrends et al., 2022). The question of between-subject variability in FC is discussed in more detail in Bijsterbosch et al. (2017).

## 2.2  Parcellations and timecourse extraction

Another important consideration when preparing fMRI data for dynamic FC analyses is the choice of parcellation and method for timecourse extraction. One may argue that, at least to some extent, the choice of parcellation depends on basic beliefs about brain organisation that have little to do with the analysis at hand. However, the way in which we divide the brain into parcels and how we extract timecourses in practice determines important aspects of the data, such as the amount of temporal and between-subject variability. Consequently, the choice of parcellation and method of timecourse extraction also greatly affect the estimation of dynamic FC (Ahrends et al., 2022; Iraji et al., 2020; Pervaiz et al., 2020). For simplicity, we will here only demonstrate the effects of two types of parcellations on temporal variability and dynamic FC: *a priori* (structural or functional) binary parcellations and data-driven functional weighted parcellations.

In a binary parcellation, the region of interest is outlined with a discrete boundary and each voxel can either belong (1) or not belong (0) to the parcel. In a data-driven functional weighted parcellation, each voxel has a specific (continuous) weight associated with each parcel. These weights are estimated from the data based on functional activity. It should also be noted that the interpretation of FC itself changes depending on how parcels are defined. For instance, if parcels are binary (meaning that each voxel can either belong to a given parcel or not) and non-overlapping (meaning that each voxel can only belong to one parcel), FC can be interpreted as connectivity between a pair of distinct regions. If, on the other hand, parcels are weighted (meaning that each voxel has a certain value in each parcel) and overlapping (meaning that each voxel belongs to some degree to several or even all parcels), FC may be interpreted as connectivity between distributed pairs of networks.

There are also several methods for extracting timecourses from voxels in these parcellations, some of which we will demonstrate below. In a binary parcellation, parcel timecourses can be extracted as the mean over voxels or the first principal component (PC). In a weighted parcellation, timecourses can be extracted using a dual regression approach to obtain individual spatial maps and individual timecourses of parcels (Beckmann et al., 2009). To understand the effects of the parcellation and timecourse extraction method on temporal variability and dynamic FC estimation, we will simulate a few different scenarios.

Let's first consider the optimal case, in which, within each parcel, voxel activity is relatively well homogenous. This is an implicit assumption of binary parcellations, but it is not usually tested explicitly. This case is illustrated in **Figure 1**. We simulate two functional clusters, *A* and *B*, each with a consistent timecourse whose peak is located at the centre of the cluster.

In this scenario, both clusters' timecourses are simply sine-waves. The only difference between the clusters is the overall signal amplitude.

```
A = zeros(10,10);
for i = 1:5
    A(i, i:(end-(i-1))) = 0.2*i;
    A(i:(end-(i-1)), i) = 0.2*i;
    A(end-(i-1), i+1:end-(i-1)) = 0.2*i;
    A(i+1:end-(i-1), end-(i-1)) = 0.2*i;
end
B = A;
T = 1:100;
yA = A(:)*sin(T);
yB = B(:)*sin(T)*0.7;
```

Here, *yA* are the timecourses for the voxels in cluster *A* and *yB* are the timecourses in cluster *B*. The voxel weights for cluster *A* and cluster *B* as well as the empirical timecourses *yA* and *yB* are shown in **Figure 1**, top row.

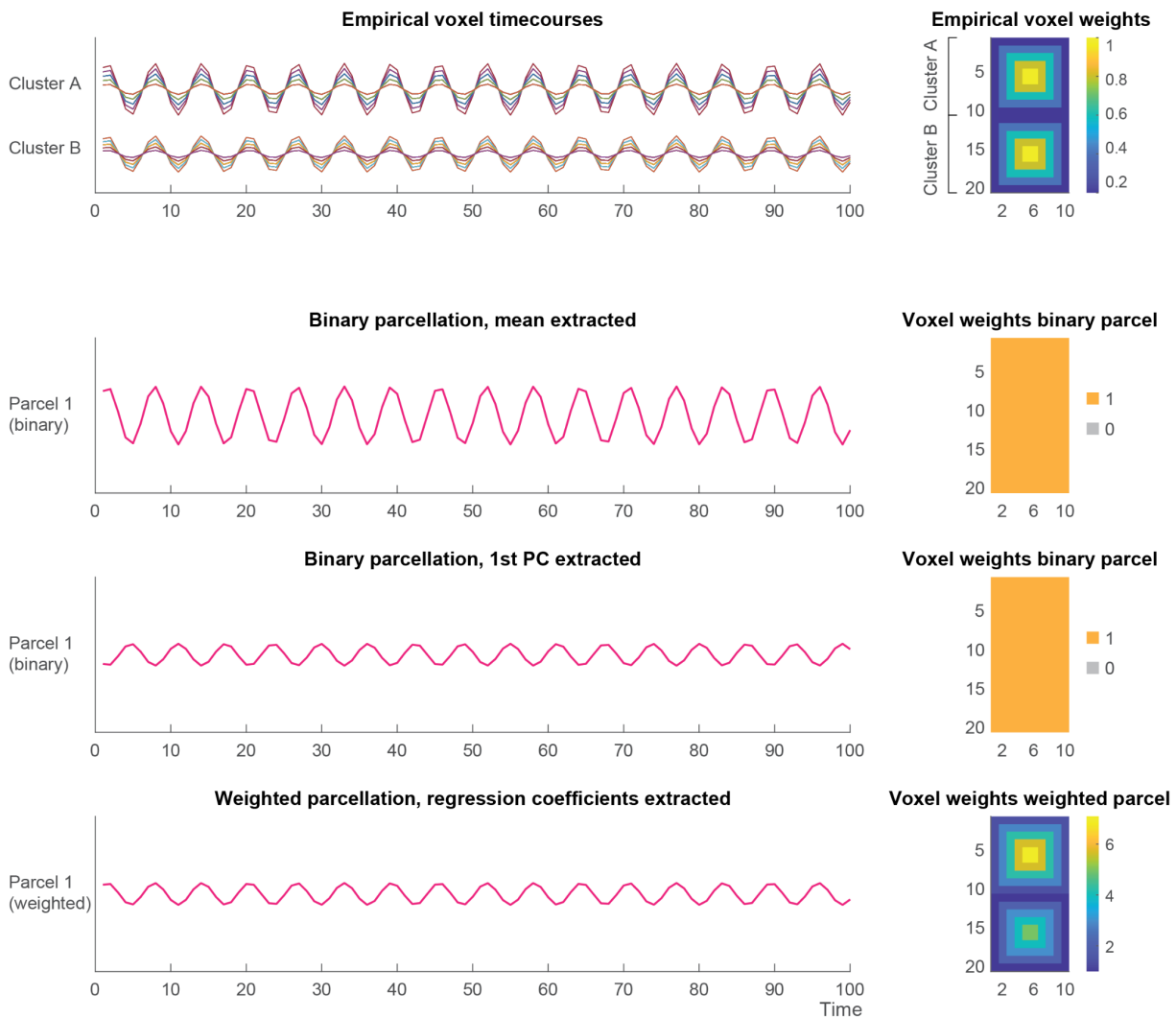

**Figure 1** *Parcellation and timecourse extraction for scenario 1, where voxel clusters in parcel have homogeneous timecourses.*

We will consider the case where both clusters are contained within the same parcel. In binary, non-overlapping parcellations, timecourses are often extracted by simply computing the mean over all voxels belonging to each parcel at each timepoint:

```
for t = T
    parcel_mean(t) = mean([yA(:,t); yB(:,t)]);
end
```

The resulting timecourses are shown in **Figure 1**, second row. In this (optimal) case, the average activity across voxels captures the overall pattern in the region adequately. However, while this method is fast and easy to compute, it can have substantial shortcomings depending on the actual temporal variance of voxel clusters in the parcel.

Consider the case where the two functional clusters' activity contained within the parcel is heterogeneous, e.g., they are negatively correlated (see **Figure 2**, top row):

```
yA = A(:)*sin(T);
yB = B(:)*-sin(T);
```

If we here extract the parcel timecourse as mean over all voxels, we will completely flatten the timecourse so that the original temporal variance will be lost (see **Figure 2**, second row). An alternative to extracting the parcel timecourse as the mean over all voxels is to use the first principal component (PC) of the voxel timecourses within the given parcel:

```
parcel_pc_tmp = pca([yA; yB]);
parcel_pc = parcel_pc_tmp(:,1);
```

The resulting timecourses are shown in **Figure 1** and **Figure 2**, third rows. Both in the case of homogeneous activity within the parcel (**Figure 1**) and in the case of heterogeneous activity within the parcel (**Figure 2**), the first PC captures the dominant pattern of voxel activity over time. In this way, we can avoid the issue of the parcel timecourse flattening out and instead preserve the most dominant temporal variance. However, using this approach, some information about the temporal variance within the region of interest would still be lost, namely that of cluster *B* in the example.

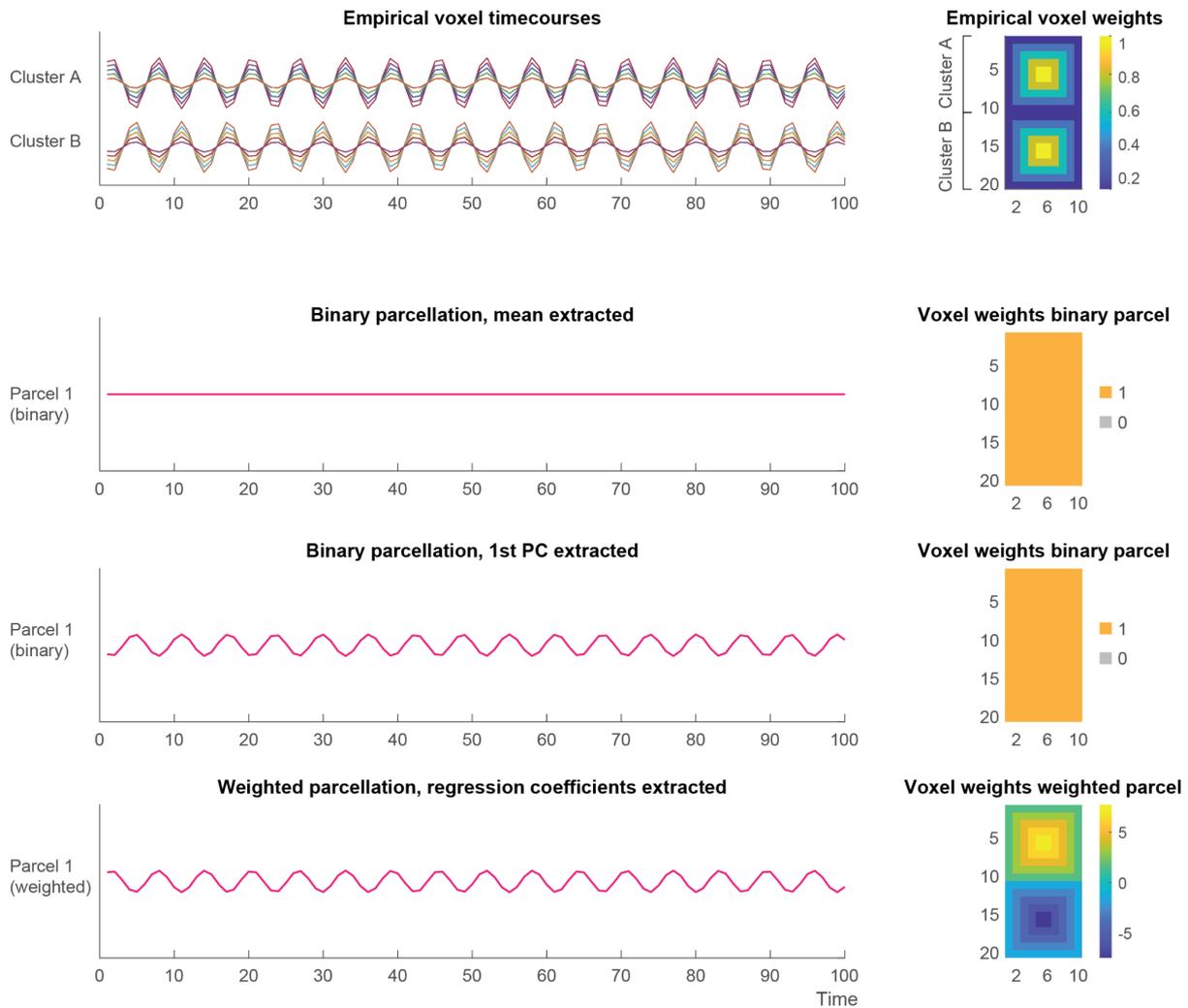

*Figure 2* Parcellations and timecourse extraction, scenario 2, where voxel clusters in parcel have heterogeneous timecourses.

Using a weighted parcellation gives a more nuanced view of the activity within a parcel. We can create a weighted parcel from the cluster timecourses by decomposing the data, for instance, using independent component analysis (ICA):

```
ica_mdl = rica([yA;yB],1);
parcel_ic = ica_mdl.TransformWeights;
```

Here, *parcel_ic* is the timecourse of a single IC extracted from the voxel-level timecourses of clusters *A* and *B*. We can access each voxel's individual weight within the parcel:

```
z = transform(ica_mdl,[yA;yB]);
weights_ICA_A = reshape(z(1:100),[10,10]);
weights_ICA_B = reshape(z(101:end),[10,10]);
```

In scenario 1 (**Figure 1**, fourth row), the IC parcel timecourse corresponds closely to the timecourse of cluster *A*. The voxel weights capture the original spatial distribution of the clusters and are higher for the voxels belonging to cluster *A* than the voxels belonging to cluster *B*. In this way, the parcel timecourse in combination with the voxel weights capture not only the most dominant temporal pattern, but also the small amplitude difference between clusters *A* and *B*. In scenario 2 (**Figure 2**, fourth row), the IC parcel timecourse again corresponds closely to the timecourse of cluster *A*. However, while the estimated voxel weights for the voxels from cluster *A* are positive, the voxels from cluster *B* have negative weights assigned, which captures how cluster *B*'s voxels' activity is negatively correlated with the activity of cluster *A* voxels. This method thus retains almost all spatiotemporal details of the original data, so that by multiplying the voxel weights with the parcel timecourses, we could almost perfectly recreate the empirical voxel timecourses.

An additional advantage of estimating the parcellation based on the functional activity from the dataset at hand is that we can avoid the issue of suboptimal parcel boundaries. This is especially important in the context of dynamic FC. Consider the case of two functional clusters, which are positively correlated at the beginning of a timeseries, then fall out of synchrony in the middle of the timeseries, and are correlated again at the end of the timeseries:

```
yA = A(:)*sin(T);
yB(:,1:30) = B(:)*sin(T(1:30));
yB(:,31:70) = B(:)*sin(31:0.5:50.5);
yB(:,71:100) = B(:)*sin(T(71:100));
```

We can calculate the average correlation between the two clusters, using e.g., a sliding window approach (see section 3.1.1 for details). This is shown in **Figure 3**, top row. Binary parcellations that are defined *a priori* on a different dataset can have the problem of suboptimal parcel boundaries. In this case, rather than dividing the data into the "true" clusters *A* and *B*, an *a priori* defined parcellation may split both clusters in the middle so that half of cluster *A* and half of cluster *B* belong to parcel 1 and the other half of each cluster belongs to parcel 2. In this case, whether timecourses are extracted using the mean or the first PC, the two clusters' activity will be contained equally in both parcels, making it impossible to disambiguate the original functional clusters.

```
for t = T
    parcel_mean(1,t) = mean([yA(1:50,t); yB(1:50,t)]);
    parcel_mean(2,t) = mean([yA(51:100,t); yB(51:100,t)]);
end
parcel_pc_tmp_1 = pca([yA(1:50,:); yB(1:50,:)]);
parcel_pc(:,1) = parcel_pc_tmp_1(:,1);
parcel_pc_tmp_2 = pca([yA(51:100,:); yB(51:100,:)]);
parcel_pc(:,2) = parcel_pc_tmp_2(:,1);
```

This is shown in **Figure 3**, second and third row for the mean timecourse extraction and first PC timecourse extraction methods, respectively. If we estimate dynamic FC between the two parcels using these timecourses, FC would be high throughout the timecourse and not capture the change in FC in the middle of the timeseries, since we are essentially computing dynamic FC of both functional clusters with themselves.

The data-driven functional parcellation (ICA) is able to capture the true parcel boundaries more accurately. We estimate ICA with two components:

```
ica_mdl = rica([yA;yB],2);
parcel_ic = ica_mdl.TransformWeights;
```

We then have a separate set of voxel weights for each component (parcel):

```
z = transform(ica_mdl,[yA;yB]);
weights_ICA_A_1 = reshape(z(1:100,1),[10,10]);
weights_ICA_B_1 = reshape(z(101:end,1),[10,10]);
weights_ICA_A_2 = reshape(z(1:100,2), [10,10]);
weights_ICA_B_2 = reshape(z(101:end,2),[10,10]);
```

These components capture the separate timecourses for clusters *A* and *B*. Estimating dynamic FC between these two components more closely resembles the empirical dynamic FC between the functional clusters (see **Figure 3**, fourth row).

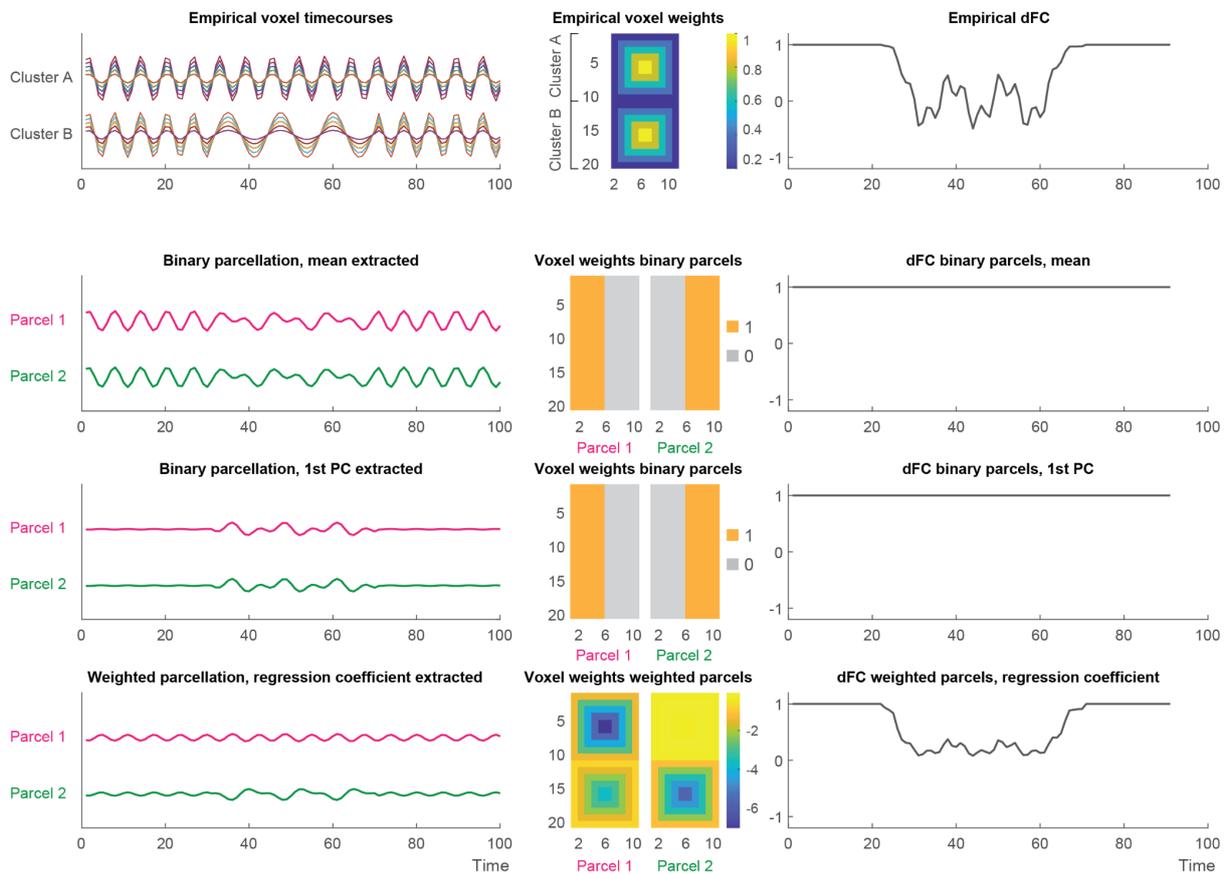

*Figure 3* Dynamic FC between two cluster timecourses (empirical: row 1) extracted from binary parcels with suboptimal parcel boundaries (row 2 and 3) vs. extracted from weighted, data-driven parcels (row 4)

The examples showed how a suboptimal parcellation or timecourse extraction method can lead to loss of temporal information and wrong estimation of the temporal relationships between parcels. When preparing data for dynamic FC analysis, it is therefore crucial to choose a suitable parcellation and timecourse extraction method. Extracting timecourses as the first PC rather than the mean can preserve temporal variability in a binary parcellation. For dynamic FC analyses, several studies found that data-driven functional parcellations, such as group ICA approaches, are preferred over binary parcellations (Ahrends et al., 2022; Iraji et al., 2020; Pervaiz et al., 2020), as data-driven parcellations not only preserve temporal variability, but also avoid the issue of suboptimal parcel boundaries that can occur with *a priori* binary parcellations. It should be mentioned that alternative approaches exist that compute time-varying FC on the voxel-level or define a parcellation based on time-varying FC (Preti & Van De Ville, 2017), but they are less common.

Beyond the type of parcellation, its coarseness needs to be taken into account. While it may be tempting to choose a parcellation to be as fine-grained as possible to maximise spatial resolution, a too fine-grained parcellation increases the number of parameters in a dynamic FC model, which can lead to noisy or even ill-posed estimations. Broadly speaking, the number of parameters to estimate in the model should be in balance with the number of observations, which are often limited in common fMRI datasets. An alternative to using a coarser parcellation is to use a dimensionality reduction technique, such as Principal Component Analysis (PCA), on the extracted timecourses of a fine-grained parcellation. However, working with transformed data in PCA-space may also introduce a bias for certain dynamic FC models (Vidaurre, 2021).

# 3   Practical Approaches to dynamic FC

There is a plethora of approaches to evaluate dynamic FC. Here, we focus on methods that fit our definition of dynamic FC, i.e., that are based on second-order statistics. This excludes, for instance, popular methods based on co-activation maps (CAPs) (X. Liu et al., 2018), which are based on first-order information. Also, we prioritise a practical understanding of the methods at the expense of breadth, so only a subset of representative methods will be discussed and analysed. For each approach there often are several software packages available.

We present methods and software to estimate single-subject, single-session dynamic FC. However, all the principles are applicable to compute the estimates over a range of sessions and subjects. This would be done by concatenating all the time series and taking into account when a session finishes and the next starts. We will assume that the data for a given scanning session have been formatted into a matrix *X*, with as many rows as fMRI volumes (i.e., time points) and as many columns as voxels, regions, or components.

## 3.1 Continuously-varying estimators

We refer as continuously-varying estimators to methods that produce an estimate of FC per time point or window, such that the parameters that define such estimations vary smoothly in the time axis. Because the dynamics are encoded by changes in these parameters, these approaches are time-varying estimates according to the above classification. We can separate two kinds of continuously-varying estimators: those based on sliding windows, which divide the data set in contiguous pieces and perform the estimation separately per piece; and all-data estimators, which use the data set all at once. Sliding windows are simpler and much more common.

### 3.1.1 Sliding windows

The most basic approach to evaluate dynamic FC is sliding windows. Here, given a certain (typically predefined) choice of window length in terms of number of frames, we perform the estimation within this window and then slide the window by one or more time points and repeat the estimation. This is done across the entire time series. For example, given *p* number of brain regions, a basic estimation would be

```
window_length = 100;
C = zeros(p,p,total_session_duration - window_length + 1);
for t = 1:total_session_duration - window_length + 1
      C(:,:,t) = corr(X(t:t+window_length-1,:));
end
```

In this code, we slide the window by one time point; for computational efficiency, the window can be slid by more than one time point. A family of variations of this scheme is related to the shape of the window: here a squared window was used, but there are alternatives that can improve the estimation in practice such as a tapered or an exponentially decaying window. In contrast to a static FC estimation that uses the entire timeseries (**Figure 4A**),

**Figure 4B** illustrates how FC is estimated by sliding windows (**Figure 4C** refers to a state-based estimation, which will be described later).

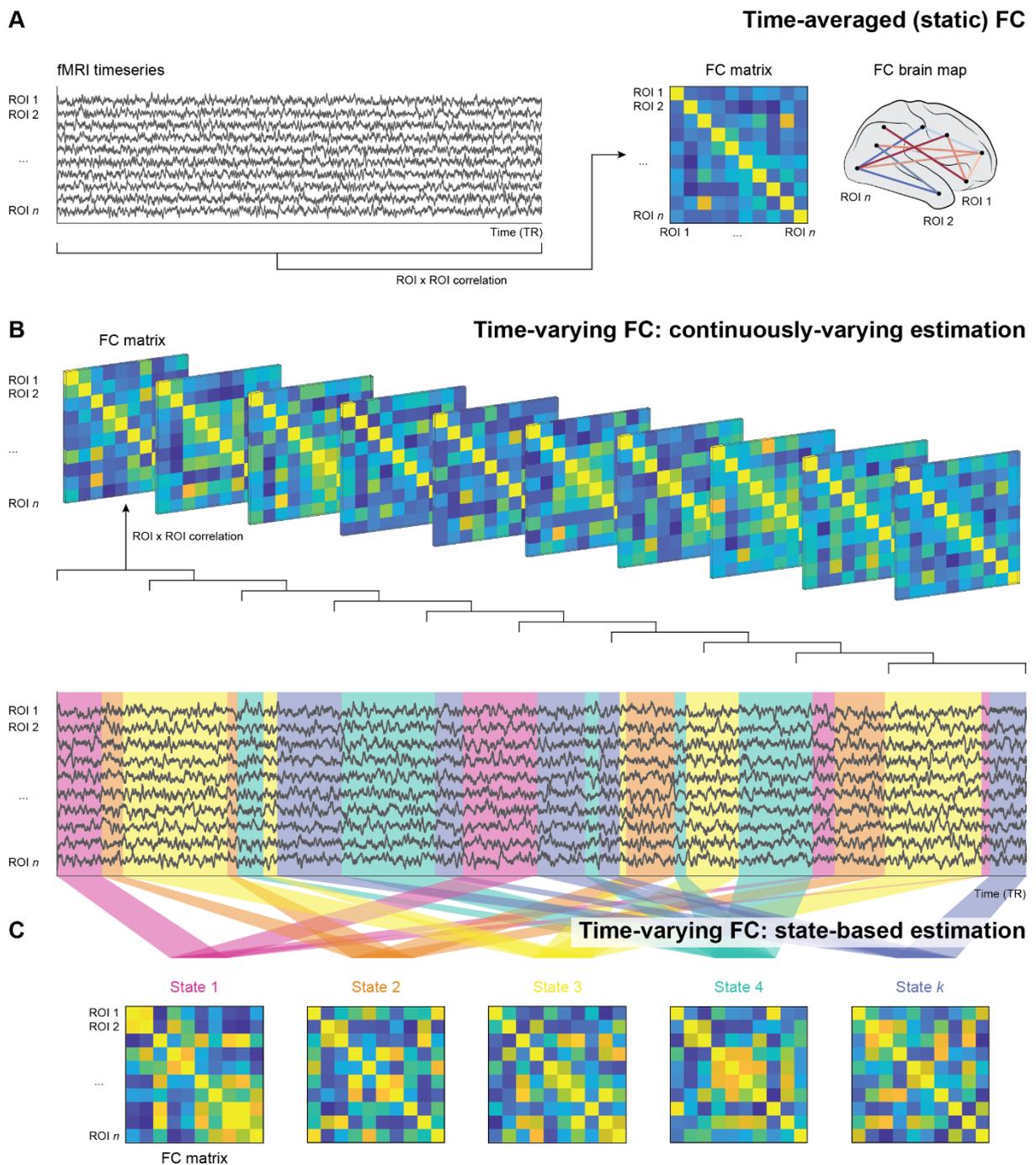

*Figure 4 Static and time-varying FC. We refer to second-order statistics of fMRI timeseries, such as correlations between regions of interest (ROIs), as functional connectivity (FC). **A)** Time-averaged (static) FC can be computed, for instance, by correlating all pairs of ROIs over the entire timeseries. The resulting FC can be illustrated as (ROI x ROI) FC matrix or as FC map in the brain. **B)** Continuously varying dynamic FC, such as sliding window approaches, estimate FC separately on portions (windows) of the timeseries. Each window has an associated FC matrix that can also be projected into the brain. **C)** State-based FC models estimate recurring patterns of FC over the timeseries. Each state has an associated FC matrix that can also be projected into the brain. (Abbreviations: ROI - region of interest; FC - functional connectivity; TR - repetition time).*

Here, we used correlations, but unnormalised covariance is also possible by using `cov()` instead of `corr()`; in this case, we will also be including information about the variance of the signal. Now, if we wish to compute partial correlation (or covariance), where the linear influence of all the other regions is removed from the coefficient between region *i* and *j*, we can consider the following code:

```
for t = 1:total_session_duration - window_length + 1
    inv_mat = inv(C(:,:,t) + lambda * eye(p));
    inv_mat = - (inv_mat ./ ...
    repmat(sqrt(abs(diag(inv_mat))),1,p)) ./ ...
        repmat(sqrt(abs(diag(inv_mat)))',p,1);
    inv_mat(inv_mat(p)>0)=0;
    iC(:,:,t) = inv_mat;
end
```

Inverse covariance matrices (here, `inv_mat`) are also referred to as precision matrices. If the number of regions is comparatively large, we need to regularise the matrix inversion to avoid badly scaled results. This is what we did here by adding the identity matrix multiplied by the regularisation constant `lambda`. Other types of regularisation that impose sparsity, by driving some partial correlations to exactly zero, are also possible. In this case, the resulting precision matrices can be mapped to a graph where non-zero coefficients relate to edges between two nodes (voxels, areas, or components), and zero coefficients relate to conditional independence between node *i* and *j*, given all the rest of the nodes (Cai et al., 2018; Friedman et al., 2008).

Sliding-window analyses can be very dependent on the choice of the window length, in the sense that too short windows will render very unstable estimates and too long windows will over-smooth the estimation and miss relatively fast changes. Although there are data-driven approaches to adaptively optimise the window length, the resulting window length is itself based on assumptions and subject to estimation noise.

A critical weakness of the sliding window approach is that, except perhaps for very long windows, a large part of the variability observed across windows will inevitably be due to estimation noise. This is because the number of time points within a window is not large enough to yield a stable measure; this gets exacerbated by the autocorrelations in the signal, which further reduce the effective degrees of freedom within the window (Afyouni et al., 2019).

This can be easily verified empirically with the following code

```
total_session_duration = 2000; p = 10;
X = mvnrnd(zeros(total_session_duration,p),true_C);
for j = 1:p
    X(:,j) = smooth(X(:,j),10);
end
X = X(101:end-100,:);
```

Here, `true_C` was a static covariance matrix obtained from real fMRI data, which we used to sample Gaussian noise. We then applied some smoothing to generate some autocorrelation in the data (imitating what we would observe in real fMRI data at a very basic level). If we then apply a sliding window analysis, we will observe broadly varying estimations of time-varying FC with either correlation or partial correlation, even though the underlying covariance matrix is not time-varying. This is illustrated in the left panel of **Figure 5A**, where we show, for a pair of signals, the results of performing a sliding window analysis on data generated as per the above code; the dotted and solid grey lines correspond, respectively, to the ground-truth correlation (i.e. `true_C(1,2)`) and empirical static correlation (i.e. `corr(X(:,1),X(:,2))`)—note the large difference between the two due to smoothing, even when using the entire length of the signals. For further discussion and examples on model validation, see **Section 4** below.

For comparison, we generated a second data set where there are two different covariance matrices underlying the generation of the data; the middle section of the time series was generated using one, and the beginning and end of the time series was generated using the other:

```
total_session_duration = 2000; p = 10;
X1a = mvnrnd(zeros(total_session_duration/4,p),true_C1);
X1b = mvnrnd(zeros(total_session_duration/4,p),true_C1);
X2 = mvnrnd(zeros(total_session_duration/2,p),true_C2);
X = [X1a; X2; X1b];
for j = 1:p
    X(:,j) = smooth(X(:,j),10);
end
X = X(101:end-100,:);
```

**Figure 5B** shows that, while the real modulation in FC is apparent, the fluctuations around it are also quite large.

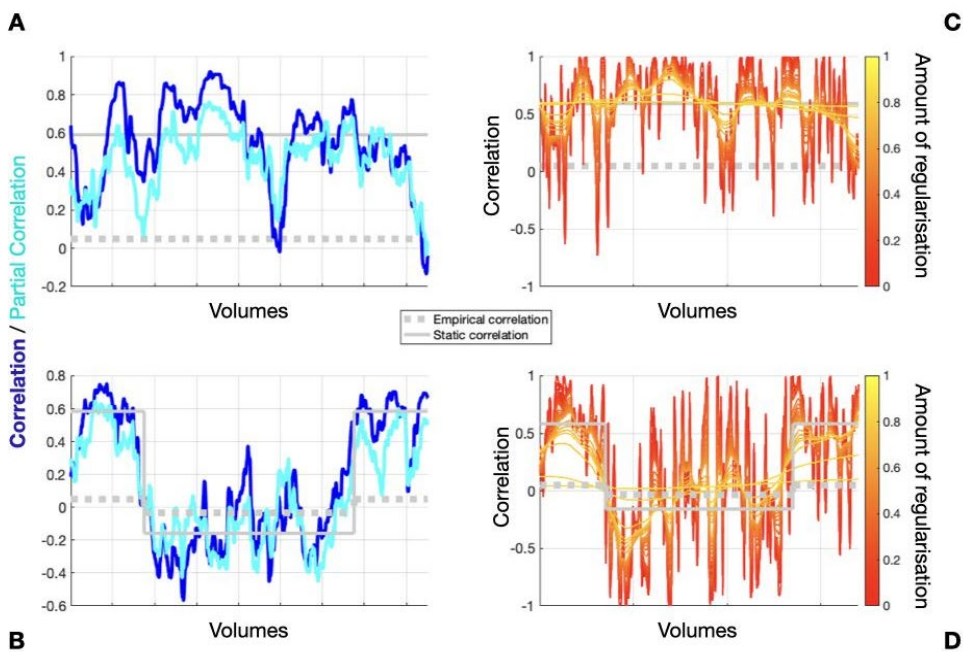

*Figure 5* Result of performing a sliding-window (**AB**) or all-data, flexible time squares (**CD**) analysis on synthetic data whose ground truth generative process either does not contain a time-varying component (**AC**), or does contain it (**CD**).

This problem has put dynamic FC on the spot of criticism. Different approaches to generate surrogate data and statistical tests have been proposed (see below), which tend to agree that sliding-window approaches struggle to find genuine time-varying FC in most data, as the variability of interest may drown in estimation noise.

### 3.1.2   All-data continuous estimators

One alternative to sliding windows within the paradigm of continuous estimators are methods that estimate one correlation coefficient per time point by considering all the data set at once. These methods do not use a predefined window, but the estimate at a given time point still depends on neighbouring time points in an adaptive fashion. This is achieved by imposing a regularisation constraint consisting of penalising the coefficient differences between contiguous time points. One such method is flexible least squares (Kalaba & Tesfatsion, 1989), which is implemented for example in the Dynamic BC toolbox (Liao et al., 2014). Multivariate approaches that estimate continuously-varying partial correlation matrices also exist based on the same idea (Monti et al., 2014). Focussing on just two signals, **Figure 5C** shows the behaviour of the flexible least squares method on the synthetic data generated above, where there was no actual fluctuation of FC in the underlying generative model of the data. As observed, choices with a low amount of regularisation (redder lines) result also in largely varying estimates (even above the 1.0 boundary), while higher amounts of regularisation converge to the empirical, non-varying correlation. When there is actual variability in FC, **Figure 5D** shows that for some choices of the regularisation parameter, the method is able to capture the modulation in the middle section of the time series above and beyond the noise.

As we can see, while continuous approaches of this kind do not necessitate the specification of a window length parameter, they are still quite dependent on the choice of the regularisation parameter. However, the fact that, unlike sliding windows, these models have a well-defined loss function lends to the use of quantitative metrics for selecting this parameter, for instance based on cross-validation. Further, as the above simulation shows, they can improve over the sliding-window estimator. On the other hand, they are more complicated to implement and they are less readily available in standard software packages.

## 3.2  State-based estimators

As opposed to the continuous estimators discussed above, state-based estimators assume that the data can be reasonably described using a discrete set of states (which can still be large). This does not mean that by using these methods we assume any one-to-one, state-to-mechanism mapping, or that there is any specific biophysical significance in the number of states; this assumption is merely descriptive. Another important assumption of these methods is exclusivity: only one state can be active at a given time.

A very important difference between continuous and state-based estimators is that, while for a continuous estimator the FC parameters are essentially all we have to estimate, for a state-based estimator we also have to estimate when the states occur (namely, the state time courses), together with each state's FC information. Another critical difference is that, because the occurrences of these states can happen anytime in the time series, a state can (and will probably) be assigned to several non-contiguous segments of the signal. Consequently, while continuous estimators (like sliding windows) do not pool information across subjects in any way, state-based estimators can have states that are shared across subjects. State-based estimators are illustrated in **Figure 4C**. Here, each of the five states' FC is estimated based on several occurrences of varying length (dwell times) throughout the timeseries.

These differences have an important practical implication: that the estimation of a state's FC is based on much more data than the estimation of a sliding window's FC. For example, let us suppose that we have 100 subjects with 20min of data per subject, and we run some state-based estimator with 20 states; each state would have on average 100 minutes of available data for the estimation of its FC parameters (e.g., a covariance or precision matrix), as opposed to a typical sliding window of 1-2min. Now, if we scale up to larger data sets like the HCP or the UK Biobank, states will have many hours of data available, effectively making the estimation of FC extremely precise except for residual states. But, as mentioned, in these methods we also have to estimate the state time courses, which is of course also subject to estimation noise. We will briefly discuss below how to characterise and deal with estimation noise at different levels.

We divide the state-based estimators into two classes: clustering methods (Allen et al., 2014) and generative models. Within the generative models, we focus on the hidden Markov models (HMM) (Vidaurre et al., 2017). Conceptually, the main difference is that the former has fewer assumptions and does not set up a model from which we can sample data. We will compare them empirically later on. There are other state-based estimators, such as the Kalman filter, that do not belong to any of these categories, but they are somewhat less common in the field of neuroimaging and will not be discussed here.

### 3.2.1 Clustering approaches

The most common clustering method builds upon sliding window estimates. Assuming we have a pool of FC matrices (one per window), the simplest procedure would be to vectorise the off-diagonal elements of the FC matrices, constructing a (no. of windows by pairs of regions) matrix, on which we will then apply a clustering technique such as k-means (Allen et al., 2014) or hierarchical clustering (Yang et al., 2014). By doing this, we would assign each window to a different state, which constitutes a categorical state time course. Probabilistic alternatives to k-means, such as a mixture of distributions, are also possible, in which case the state time courses will be made of probabilities (that is, per window and state). Together with the estimated state time courses, the state FC estimations would correspond to the centre of the clusters. The hope of this method is that, even though the individual sliding window estimates are quite noisy, the state FC matrices are made of averaging across many windows and therefore will be less noisy.

A mathematically more principled version of this approach is based on the use of Riemannian distances between the window FC matrices (instead of Euclidean, as results from vectorising the FC matrices) (Pervaiz et al., 2020).

For simplicity, we will here illustrate only the simplest (and most common) approach with code. Assuming we have already computed sliding-window estimates on data generated using the code above, the following code implements the most basic clustering approach:

```
N_windows = total_session_duration - window_length + 1;
C_unwrapped = zeros(N_windows,p*(p-1)/2);
for j = 1:N_windows
    Cj = C(:,:,j); Cj = Cj(triu(true(p),1));
    C_unwrapped(j,:) = Cj(:);
end
K = 4;
idx = kmeans(C_unwrapped,K);
state_time_courses = zeros(length(T),K);
for k = 1:K
   state_time_courses(idx==k,k) = 1;
end
```

We can now get the fractional occupancies (i.e. the proportion of time taken by each state), and display the state time courses as

```
mean(state_time_courses)
area(state_time_courses)
```

Another clustering approach that does not necessitate the specification of a window length hyperparameter is Leading Eigenvector Dynamics Analysis (LEiDA) (Cabral et al., 2017). In short, LEiDA computes functional connectivity estimates at each time point by first computing the signals' instantaneous phase (by means of the Hilbert transform) and then calculating the cosine similarity between each pair of signals. Note that even though phase is instantaneous, it still needs information from neighbouring time points. This yields a relatively noisy estimation of functional connectivity per time point. To reduce the amount of noise, the first eigenvector of each FC matrix is extracted using a singular value decomposition, producing a (no. of time points by no. of regions) matrix containing FC information. We can then apply a k-means algorithm as usual. For illustration, this procedure is implemented in the code below:

```
Phase = zeros(total_session_duration,p);
for j=1:p
        Phase(:,j) = angle(hilbert(X(:,j)));
end
Eigenvectors = zeros(total_session_duration,p);
for t = 1:total_session_duration
        C = zeros(p);
        for j1 = 1:p-1
                for j2 = j1+1:p
                        d = abs(Phase(t,j1) - Phase(t,j2));
                        if d > pi, d = 2*pi-d; end
                        C(j1,j2) = cos(d);
                        C(j2,j1) = C(j1,j2);
                end
        end
        [v,d] = eig(C);
        [~,i] = max(diag(d));
        Eigenvectors(t,:) = v(:,i);
end
[idx,eigen_centroids] = kmeans(Eigenvectors,K);
```

Here, `eigen_centroids` define the states. Each state essentially projects the brain areas into a one-dimensional axis, such that areas close to each other in one extreme of the axis (i.e., having the same sign in `eigen_centroids`) are in-phase, and areas that belong to opposite extremes of the axis (i.e., having opposite signs) are anti-phase. Note that the sign in `eigen_centroids` is arbitrary, and only the sign relationships between areas are meaningful.

Overall, clustering approaches are generally simple to implement but they are not generative models, i.e., we do not have a compact set of parameters to sample data or perform statistical testing. The hidden Markov model (HMM), discussed in the next section, is a generative model.

### 3.2.2 Generative models: Hidden Markov Models (HMMs)

Generative models are those that specify a full mathematical structure from which we can sample new data sets (Bishop, 2006). As such, they define a probability distribution over the data, and have a well-defined set of parameters over which we can do statistical inference and testing. Here, we discuss the HMM, a generative model specifically designed to deal with temporal (and other sequential) data (Vidaurre et al., 2018; Vidaurre et al., 2017).

The HMM is, rather than a single model, a family of probability distributions where each state is in itself a probability distribution, defined by a set of state parameters. By choosing one or another state distribution (also called observation models), the HMM can adapt to different kinds of data. For example, state distributions can be Gaussian, multinomial, Wishart, Poisson, etc. Here, we will focus on the Gaussian distribution because it is the common choice for fMRI data. A Gaussian distribution is parametrised by a mean vector and a covariance matrix containing FC information. This allows three main variants of the Gaussian HMM. First, having a shared covariance matrix for all states and one mean vector per state; this configuration however does not model changes in FC and will not be discussed here. Second, we can have one mean vector and covariance matrix per state, which is the most common approach. Third, in order to focus on FC, we can pin the state means to zero and have only a covariance matrix per state (which is equivalent to having a Wishart state distribution) (Vidaurre et al., 2021).

The remaining parameters of the HMM are the initial probabilities, i.e., the probability of the trials to commence with a given state; and the transition probability matrix, with elements ($j$,$k$) encoding the probability of transitioning from state $j$ to state $k$. The initial probabilities, as well as each of the rows of the transition probability matrix, are modelled as Dirichlet distributions.

Apart from state exclusivity, and the fact that we use a discrete number of states, the use of a transition probability matrix implies that the HMM builds upon a third assumption: Markovianity. In the temporal domain, this means that which state is active in the present time point depends on which state was active in the previous time point; or in more rigorous terms, that the state at time point $t$ is conditional independent to all the rest of the time points given $t$-1 and $t$+1. In practice, this typically leads to having smoother, and not too abrupt, state transitions. Note that even though the HMM does not model longer term dependencies, the model does not forbid them either, and the resulting state time courses might still exhibit long-term dependencies (Vidaurre et al., 2017). Versions of the HMM that explicitly model the state visit durations, therefore dealing with longer term dependencies, are referred to as semi-Markovian (HsMM). These have also been applied to fMRI data, in practice yielding similar estimations to the standard HMM (Shappell et al., 2019).

Now, the inference of an HMM from data implies not only the estimation of the HMM parameters just described but also the state time courses. While the HMM model can be ported to different data sets, the state time courses are specific to the data set on which they were estimated. To illustrate the use of the HMM and its inference, we will use the HMM-MAR[1] toolbox in Matlab. Assuming that we have added the toolbox to the Matlab path, we can estimate a model by

```
options = struct();
options.K = 8;
options.covtype = 'full';
[hmm,Gamma,~,vpath] = hmmmar(X,T,options);
```

With this code, we are estimating an HMM model with K=8 states, and state time courses Gamma, on data X, which might be composed of different concatenated fMRI sessions or subjects. The length of each session is indicated in vector T (in number of time points). If we wish to pin the mean of the state Gaussian distributions to zero such that the states are defined only as FC matrices, we would use options.zeromean = 1.

---

[1] The name of the toolbox has a historical reason: the first observation model available was the multivariate regressive model. But it now contains other distributions, such as Gaussian, Poisson, Wishart, probabilistic PCA, or the multivariate regression model. All documentation is available at https://github.com/OHBA-analysis/HMM-MAR/wiki

Using the same variables for `T` and `options` that we used for training, we can query the fractional occupancies (the proportion of activation per state and subject) and plot the state time courses as

```
FO = getFractionalOccupancy(Gamma,T,options)
for k = 1:8
    state_time_courses(vpath==k,k) = 1;
end
figure; imagesc(FO)
figure; area(state_time_courses(1:T(1),:))
```

For interpretation purposes, it is important to note that, unlike the sliding window estimates (which are typically correlations matrices), here a FC matrix is a covariance matrix and therefore also conveys information about the variance of the signal. As mentioned above, although the variance of an entire scanning session is one because of standardisation, the variance for sub-periods of the signal might not be exactly one. As a consequence, the HMM could be capturing changes in amplitude or variance in certain cases.

A final remark is about the estimation of the number of states. This could be done in different ways, for example using the free energy. In real data this question is not very relevant because the ground-truth number of states does not exist; more states will simply result in a finer-grained estimation, and replicability is often a more sensible criterion (Vidaurre et al., 2017).

## 3.3  Multivariate autoregressive models

So far, we have been considering dynamic functional connectivity in terms of time-varying instantaneous FC. In this section, we will describe the multivariate autoregressive model (MAR), a linear dynamical system that models non-instantaneous aspects of FC with a single set of parameters —i.e., without the use of states or windows (Harrison et al., 2003; Rogers et al., 2010).

Mathematically, we can define the MAR model as the following generative model:

$$X_t = \sum_{j=1}^{L} X_{t-j} A^j + \varepsilon_t$$

Where $\varepsilon_t$ is Gaussian noise, $A_j$ are matrices of autoregressive parameters and $L$ is the order of the MAR model, which defines the complexity of the model and which we have to choose beforehand. Order selection can make use of a pre-specified criterion, such Akaike's (Akaike, 1974), or use cross-validation; we will however not discuss this here since, in practice, a precise estimation of the optimal order is not that relevant in neuroimaging. The MAR model can be expanded with non-linear terms (e.g., area interactions), but we will not discuss this possibility here either. We can reexpress the above equation in a standard regression form as

$$X_t = Y_t A^* + \varepsilon_t$$

where $Y_t$ is defined as $[X_{t-L} \dots X_t]$ and $A^*$ is a ($L\,p$ by $p$) matrix that reunites all the $A_j$ matrices across the $p$ channels. This allows us to straightforwardly apply any standard regression estimation method to find the autoregressive parameters. For example, assuming that there is only one subject in $X$, we can compute the autoregressive coefficients using least squares:

```
Y = zeros(size(X,1) - L , L * p);
for j = 1:L
        Y(:,(1:p)+(j-1)*p) = X(L-j+1:end-j,:);
end
A_star = pinv(Y) * X(L+1:end,:);
```

Although very simple to estimate, the MAR model contains rich information about the signal on the frequency domain. Power, coherence, partial directed coherence, etc. can be readily computed by Fourier transforming $A^*$ (in the code, A_star), and the richness of these spectral estimates depends on the order of the model. This falls out of the scope of this chapter; but see Faes and Nollo (2011) for a comprehensive account.

According to these equations, the diagonal elements of the $A^j$ matrices (and the corresponding elements in $A^*$) refer to "self-connections", and the off-diagonal elements refer to cross-area connections, which we can interpret as dynamic FC. Note that, similarly to the elements in a covariance matrix, the autoregressive coefficients are second-order statistics, but are not instantaneous.

Conveniently, the autoregressive coefficients share with the precision matrix the property that, for a very large amount of data, the coefficient(s) between variable *i* and variable *j* will be exactly zero if they do not hold a direct linear (in this case lagged) relationship, i.e., after we regress out the other variables. This has opened the window for the MAR model to be used as a tool to infer effective connectivity. Critically, this is under the assumption that there are no unobserved variables in the system. For example, we could get a non-zero coefficient between two variables if these were connected through an unobserved third variable but not directly. Of course, this assumption does not often hold in real neuroimaging data, but this property could still be useful because, at least, if a coefficient is zero then we can assert that the variables are not linearly connected.

Let us show the most basic scenario with an example, where we assume the following linear dynamical system of order 1 with variables (brain areas) $X^{(1)},..., X^{(5)}$:

$$X^{(1)}{}_t = 0.5\,X^{(1)}{}_{t-1} + 0.25\,X^{(2)}{}_{t-1} + \varepsilon^{(1)}{}_t$$

$$X^{(2)}{}_t = 0.5\,X^{(2)}{}_{t-1} + 0.25\,X^{(1)}{}_{t-1} + 0.25\,X^{(3)}{}_{t-1} + \varepsilon^{(2)}{}_t$$

$$X^{(3)}{}_t = 0.5\,X^{(3)}{}_{t-1} + 0.25\,X^{(2)}{}_{t-1} + 0.25\,X^{(4)}{}_{t-1} + \varepsilon^{(3)}{}_t$$

$$X^{(4)}{}_t = 0.5\,X^{(4)}{}_{t-1} + 0.25\,X^{(3)}{}_{t-1} + 0.25\,X^{(5)}{}_{t-1} + \varepsilon^{(4)}{}_t$$

$$X^{(5)}{}_t = 0.5\,X^{(5)}{}_{t-1} + 0.25\,X^{(4)}{}_{t-1} + \varepsilon^{(5)}{}_t$$

And let us assume we sample data from this system with the following code:

```
p = 5;
total_session_duration = 1000;
true_A = 0.5*eye(p);
for j = 1:p-1
    true_A(j,j+1) = 0.25; true_A(j+1,j) = 0.25;
end
X = randn(total_session_duration,p);
for t = 2:total_session_duration
    X(t,:) = X(t,:) + X(t-1,:) * true_A;
end
```

Now, if we sample many data sets like this and fit autoregressive models with the code above, we would obtain distributions of estimated parameters as in **Figure 6**. If instead of being able to sample many data sets, we had a limited number of subjects, we could still perform a bootstrapped estimation (where we create pseudo data sets by randomly sampling subjects with repetitions).

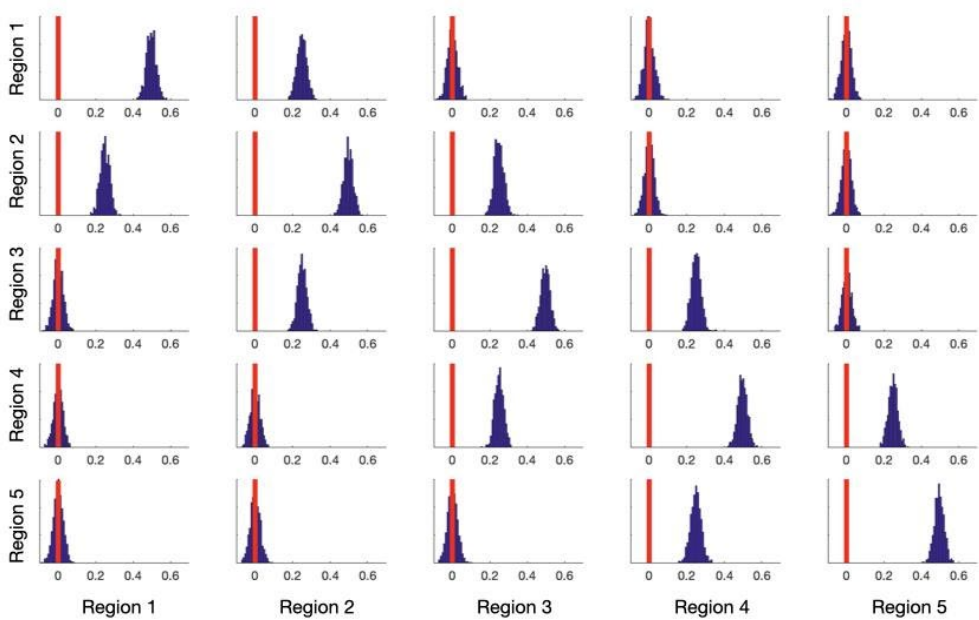

*Figure 6* Estimation of autoregressive coefficients over 1000 data sets generated through an idealised linear dynamical system.

Of course, this is an extremely idealised scenario. In practice, the ground-truth is not an autoregressive model, and autoregressive coefficients are rarely zero in real data. The concept of Granger causality emerged as a dedicated statistical test in the field of economics to deal with real data under a definition of causality based on temporal precedence.

However, the use of Granger causality in fMRI data for inferring effective connectivity has been criticised because of the differences in latency in the hemodynamic response function across brain regions, as well as other reasons.

Still, multivariate autoregressive modelling is a useful FC statistic in fMRI (as opposed to an effective connectivity statistic), because it is easy to implement, it is relatively light in assumptions while still informative, and, used as a predictor of behavioural traits, it can surpass the accuracy of static FC estimates (Liégeois et al., 2019).

# 4 Evaluating dynamic FC models

In the previous section, we outlined some representative methods for dynamic FC estimation. In this section, we discuss how to validate these models.

## 4.1 Testing against null models

A prevalent discussion in dynamic FC research (particularly in the sense of time-varying) is the question whether FC meaningfully fluctuates over time, or whether it is actually stable over time and temporal fluctuations are mainly due to noise. Depending on the criteria, it has both been argued that real fMRI data fluctuates over time (Zalesky et al., 2014) and that it is difficult to reject the statistical hypothesis that FC is stationary (Hindriks et al., 2016; Liégeois et al., 2017; Lindquist et al., 2014). An argument for the latter view is that what may be interpreted as dynamic FC can in fact be explained, for instance, by sampling variability (Laumann et al., 2017), which may in turn be affected by denoising strategies. Simulation studies also address whether static FC is being driven by transient (dynamic) events (Zamani Esfahlani et al., 2020) or whether the presence of transient events in FC may be driven by static FC (Ladwig et al., 2022; Novelli & Razi, 2022). Although some time-varying FC methods are in principle equipped to find when the data is not time-varying (for example, in a state-based model by removing all states but one), in practice they may find temporal variation in FC due to statistical noise and sampling variability even if FC in the data was really stationary.

We can see this effect of sampling variability by simulating data where FC does not vary over time, using real data as a starting point. In this case, a null model is a model that assumes stationarity of FC while preserving as many of the general distribution properties of real fMRI data. We can, for example, generate data from a multivariate Gaussian distribution that preserves the mean and variance of the real fMRI data:

```
T = 1000;
n_areas = size(tc_real,2);
mu = mean(tc_real);
Sigma = cov(tc_real);
tc_mvn = mvnrnd(mu, Sigma,T);
```

We can now compare the sliding window correlations between the real data and the simulated data (for simplicity, we only look at the correlation between the first two areas):

```
window_length = 10;
real_dFC_tmp = zeros(n_areas, n_areas, T - window_length + 1);
mvn_dFC_tmp = zeros(n_areas, n_areas, T - window_length + 1);
for t = 1:T - window_length + 1
    real_dFC_tmp(:,:,t) = corr(tc_real(t:t+window_length-1,:));
    real_dFC(1,t) = real_dFC_tmp(1,2,t);
    mvn_dFC_tmp(:,:,t) = corr(tc_mvn(t:t+window_length-1,:));
    mvn_dFC(1,t) = mvn_dFC_tmp(1,2,t);
end
```

As we show in **Figure 7**, for a sufficiently small window size, the sliding-window estimation of FC in the surrogate data will vary considerably over time (see **Figure 7B, middle panel**; and see also **Figure 5**), even though the true covariance of the data is static. We can increase the window to see if we can recover the true (static) nature of the correlation:

```
window_length = 200;
real_dFC_tmp = zeros(n_areas, n_areas, T - window_length + 1);
mvn_dFC_tmp = zeros(n_areas, n_areas, T - window_length + 1);
for t = 1:T - window_length + 1
    real_dFC_tmp(:,:,t) = corr(tc_real(t:t+window_length-1,:));
    real_dFC(1,t) = real_dFC_tmp(1,2,t);
    mvn_dFC_tmp(:,:,t) = corr(tc_mvn(t:t+window_length-1,:));
    mvn_dFC(1,t) = mvn_dFC_tmp(1,2,t);
end
```

As we increase the window size, the correlation in the surrogate data flattens out (see **Figure 7B**, bottom panel), while we can still see some temporal variation in the sliding window correlations of the real data (see **Figure 7A**, bottom panel). Here, it may seem as though FC in the real fMRI data varies more over time than FC in the surrogate data. To test this hypothesis, we would generate many random time courses from the multivariate Gaussian distribution and test for significant differences in variance between the real fMRI data and the surrogate data. However, multivariate Gaussian null data is not a very appropriate null model in this case, since it does not capture any of the autocorrelations present in real fMRI data.

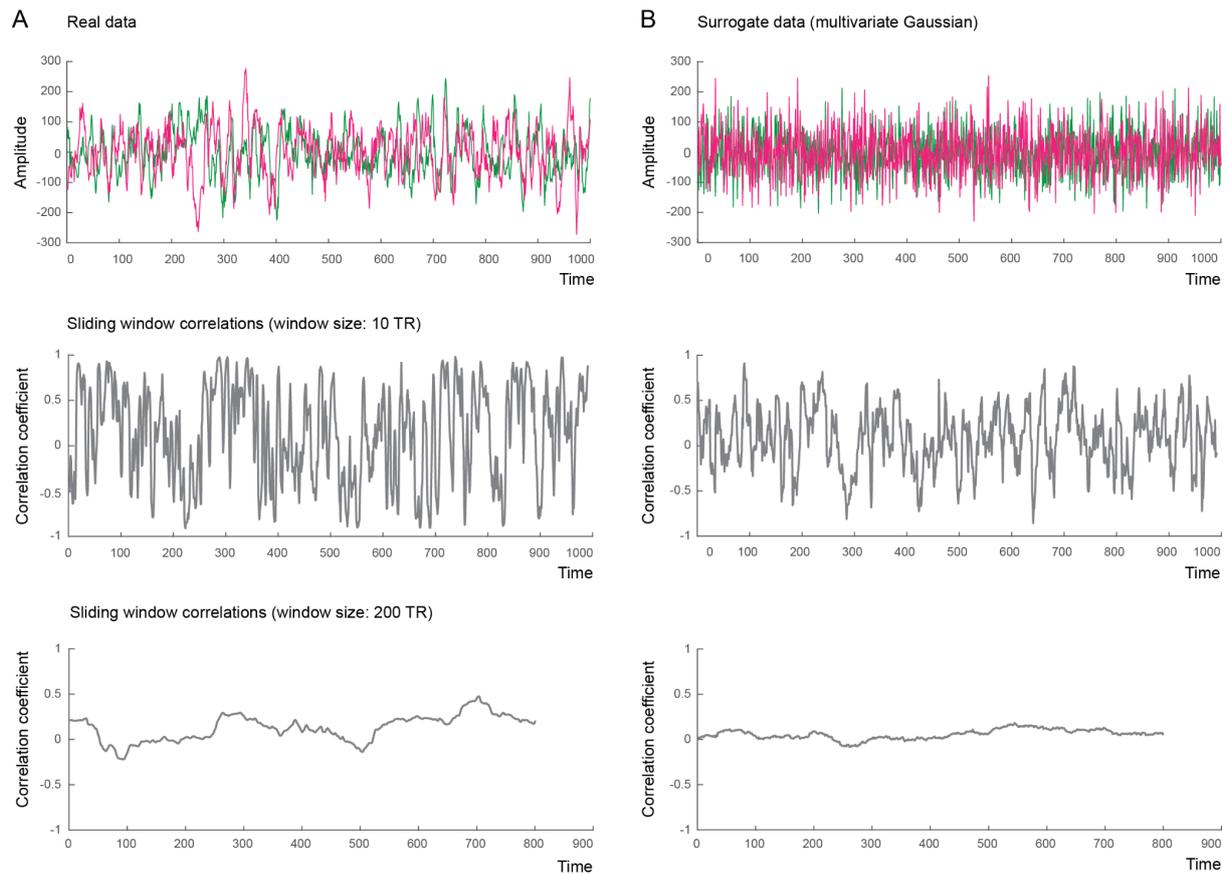

**Figure 7** *Is FC actually dynamic or are temporally changing estimates of FC a result of sampling variability?* **A)** *Real fMRI timecourses for two regions (top row) and their corresponding sliding window correlations at a window size of 10 TRs (middle row) and 200 TRs (bottom row).* **B)** *Surrogate timecourses for two regions, randomly generated from a multivariate Gaussian distribution with the same mean and covariance as the real fMRI data (top row), and their corresponding sliding window correlations at a window size of 10 TRs (middle row) and 200 TRs (bottom row).*

An important step in the development and evaluation of dynamic FC methods is therefore to have *suitable* null models. However, defining the distribution of real fMRI data and setting criteria for stationarity is not straightforward.

As a more adequate alternative to multivariate Gaussian models, we can use null models that preserve not only the mean and variance, but also the autocorrelation structure of the real fMRI data. We will demonstrate this following the example in Liégeois et al. (2017) to generate surrogate data from a first order autoregressive (AR-1) model. To do this, we first estimate the 1-lag autoregressive distribution parameters of the real fMRI data:

```
Y = tc_real(2:end,:)';
X(1,:) = ones(1,T-1);
X(2:n_areas+1,:) = tc_real(1:T-1, :)';
beta = (Y*X')/(X*X');
residuals = Y-beta*X;
```

We then generate synthetic data with these distribution parameters:

```
c = beta(:,1);
weights = beta(:, 2:n_areas+1);
tc_real_flip = tc_real';
tc_arr = zeros(T, n_areas);
tc_arr(1,:) = tc_real_flip(:,randi(T-1))';
rand_t = randperm(T-1);
res_mu = mean(residuals,2);
res_Sigma = cov(residuals');
noise = (mvnrnd(res_mu,res_Sigma,T-1))';
for i = 2:T
    tc_arr(i,:) = c' + (weights*tc_arr(i-1,:)')' + ...
        noise(:,rand_t(i-1))';
end
```

Finally, we again compute the sliding window correlations in the surrogate timeseries and compare them to the sliding window correlations of the real fMRI data we computed earlier:

```
window_length = 200;
arr_dFC_tmp = zeros(n_areas, n_areas, T - window_length + 1);
for t = 1:T - window_length + 1
    arr_dFC_tmp(:,:,t) = corr(tc_arr(t:t + window_length-1,:));
    arr_dFC(1,t) = arr_dFC_tmp(1,2,t);
end
```

The resulting surrogate timeseries and corresponding sliding window correlations are shown in **Figure 8B**. Here, the sliding window correlations of the surrogate data have similar temporal variance to the real fMRI data.

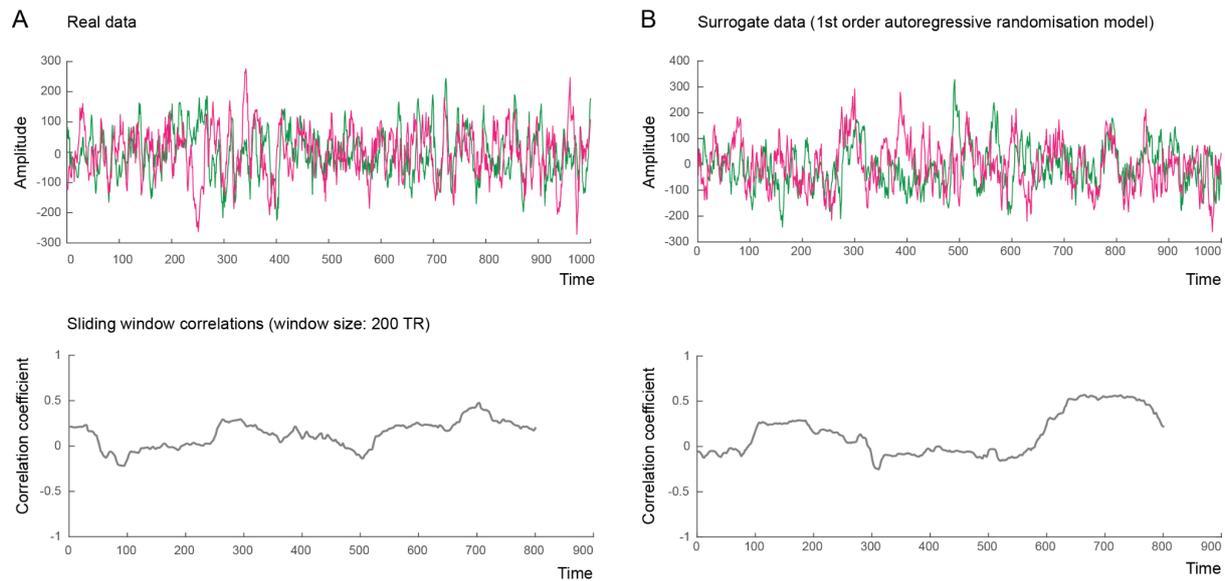

*Figure 8 When generating surrogate data from a model that preserves the autocorrelation structure of the real fMRI data, such as a 1-lag autoregressive randomisation model, the null hypothesis cannot easily be rejected.*

In fact, several studies (Hindriks et al., 2016; Liégeois et al., 2017; Prichard & Theiler, 1994) showed that for real fMRI data, the null hypothesis often cannot be rejected when using surrogate data that preserves the autocorrelation structure of the real data, such as ARR models or models based on phase randomisation (PR). The difference between the ARR model and the PR model is how much of the autocovariance structure they preserve (the PR model preserves the entire autocovariance structure, while the $p$-th order ARR model preserves only the first $p + 1$ autocovariances).

It is however important to note that, since the assumption of both the AR model and the PR model is that fMRI data are realisations of a linear, Gaussian, and stationary process, a rejection of the null hypothesis in this case could be due to any of these assumptions being untrue (i.e., not necessarily stationarity). That is, not being able to reject the null hypothesis does not necessarily imply that the data is stationary. Liégeois et al. (2017) also showed that the opposite conclusion also does not hold, i.e., being able to reject the null hypothesis does not necessarily imply that the data is dynamic. While testing fMRI data for non-stationarity is therefore not trivial, tests against null models can be useful to understand which information different models use in their estimation of FC dynamics (Liégeois et al., 2021).

## 4.2  Comparing models with simulated data

An important question is how we can test the advantages and disadvantages of the different methods. A powerful way to do this is through the use of synthetic data, where we can know and control the ground-truth generative model. There are many ways to generate synthetic data; here, we discuss as an illustration a simple procedure to compare the basic clustering approach to the HMM. The idea is to use real fMRI data as a starting point. This approach is described in detail in Ahrends et al. (2022). For example, let us assume that we have data from ten regions in the brain from one subject. We can compute the covariance matrix of those real data and generate data from a Gaussian distribution, which we can smooth to make it look a bit more like real fMRI data:

```
some_time = 5000;
p = 10;
C = cov(real_data);
X = mvnrnd(zeros(some_time,p),C);
for j = 1:p
    X(:,j) = smooth(X(:,j),10);
end
```

Now, from this empirical covariance matrix we can generate covariance matrices with small variations. For this, we can perform a singular value decomposition of the empirical covariance matrix, permute the low-order eigenvectors, and reassemble the matrix. Depending on how many eigenvectors we permute (in the following code the last `J` eigenvectors) and their corresponding eigenvalues, we can make the variations larger or smaller:

```
[U,S,~] = svd(C);
e = diag(S); e = cumsum(e) / sum(e);
C_synth = zeros(size(C));
% less than 90% variation
J = find(e>=0.9,1);
for j = 1:J
        C_synth = C_synth + U(:,j) * S(j,j) * U(:,j)';
end
for j = J+1:10
        c = U(randperm(size(U,1)),j);
        C_synth = C_synth + c * S(j,j) * c';
end
```

We can use this code to generate a set of states, each with a different (but still relatively similar) covariance matrix. This way, the larger the variations (i.e., the largest is the sum of the eigenvalues of the permuted eigenvectors), the stronger the between-state differences and the easier it is to find the ground-truth states by the HMM or any other method.

We can then sample ground-truth state time courses, for four states for example, as

```
total_session_duration = 20000;
max_duration = 100; min_duration = 5;
true_stc = zeros(total_session_duration,1);
t=1;
while t <= total_session_duration
        k = randi(4,1);
        L = randi(max_duration-min_duration,1) + min_duration;
        if t+L-1 > total_session_duration
                L = total_session_duration - t + 1;
        end
        true_stc(t:t+L-1,k) = 1;
        t = t + L + 1;
end
```

In this case, we randomly sampled the identity of the state from a categorical distribution and the duration of the state visits uniformly from the interval [min_duration,max_duration], which corresponds to a form of semi-Markov process. Many other sampling schemes are possible, with which we can interrogate the practical relevance of the models' assumptions (for instance, the HMM's Markovianity; see below).

Altogether, we can put these different pieces together to sample synthetic data sets: sample of covariance matrices, sample of state time courses, sample of data, and final smoothing. Next, we will use simulated data from this scheme to compare the basic clustering approach and the HMM. We followed these steps to generate 200 data sets for each of six different scenarios:

- Slower transitions (`max_duration = 1000; min_duration = 50`) and larger state variations (`J = 1`).

- Faster transitions (`max_duration = 100; min_duration = 5`) and larger state variations (`J = 1`).

- Slower transitions (`max_duration = 1000; min_duration = 50`) and moderate state variations (`J = 3`).

- Faster transitions (`max_duration = 100; min_duration = 5`) and moderate state variations (`J = 3`).

- Slower transitions (`max_duration = 1000; min_duration = 50`) and smaller state variations (`J = 6`).

- Faster transitions (`max_duration = 100; min_duration = 5`) and smaller state variations (`J = 6`).

On each of these 4 x 200 data sets, we ran the basic clustering approach and the HMM, both endowed with four states, matching the ground truth. Each method produced an estimated state time course. Since the order of the states is not identifiable (i.e., the first state in the ground-truth state time courses could correspond to the third state of the estimated state time courses), we used the Hungarian algorithm (Kuhn, 1955) to align them (as implemented in the HMM-MAR toolbox). We then computed the accuracy as the correlation of the state probabilities with the ground-truth state time courses

```
options = struct();
options.K = 4;
options.covtype = 'full';
options.zeromean = 1; % states only have a covariance matrix
[hmm,stc] = hmmmar(X,size(X,1),options);
[assig,cost] = munkres(1-corr(true_stc,stc));
stc = stc(:,assig);
accuracy = corr(stc(:),true_stc(:));
```

**Figure 9** shows the accuracy of each run per method and scenario (dots where a jitter was introduced in the x-axis for ease of visualisation), together with the across-run average (bars). As observed, the HMM fared better than the basic clustering approach for two reasons. First, because the estimation of the states for the clustering method is based on noisy sliding window estimates, whereas the HMM estimates the states directly from the data using a much larger number of time points and, therefore, does it so much more precisely; for this reason, the difference between the HMM and the clustering approach is considerably smaller when the state covariance matrices are very different between states (upper row of panels). Second, because the clustering approach cannot detect fast changes (i.e., faster than the length of the window), whereas the HMM states can switch as quickly as necessary; that is why the accuracies for clustering approach are lower in the right panels, whereas the speed of the ground-truth state transitions makes less of a difference for the HMM.

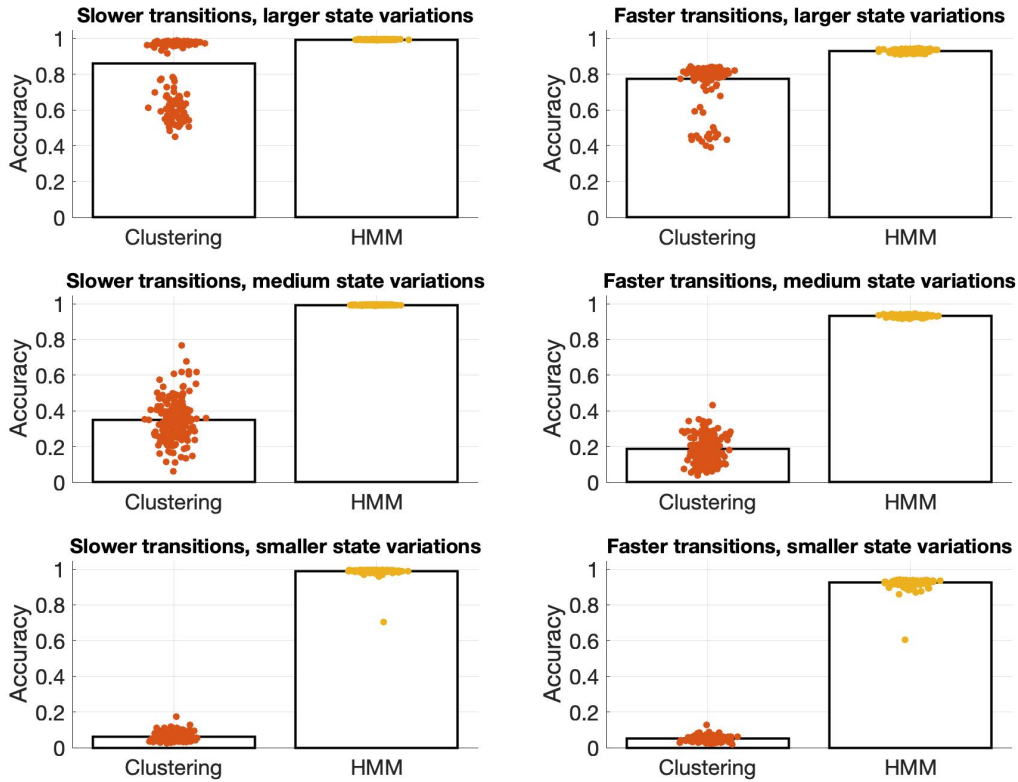

**Figure 9** *Performance of the standard clustering approach vs. the HMM in estimating the state transitions sampled from a state-based generative model where the state distributions are covariance matrices. Both how different the states are, and the speed of the transitions are manipulated to create simpler or harder estimation problems. Each dot represents a run (a sampled data set and the corresponding estimates) and the bars are averages.*

Although not shown here, LEiDA performed notably worse than the other methods. The reason is that, while LEiDA's states are based on the first principal component of matrices of phase relationships, the states of both the HMM and the clustering approach are actual covariance matrices, matching the ground-truth state model. This is a general principle of this type of simulations: the more accurate the assumptions of a method are with respect to the ground-truth generative process, the more accurate its estimates will be. It is important to note that, in real data, where the generator of the data is the brain, LEiDA will just offer a different perspective: in this case, we can hardly say that one model is more correct than another.

# 5 Challenges & perspectives

Having discussed the most important practical aspects of conducting dynamic FC analyses, we will now outline some challenges and perspectives in the current development of dynamic FC. Challenges and questions in the dynamic FC research field are also extensively discussed in Lurie et al. (2019), so we here only highlight a few key ideas related to reliability and reproducibility, and the use of dynamic FC to predict behavioural and clinical traits.

## 5.1 Reliability & reproducibility

A key challenge in the study of dynamic FC is the reliability and reproducibility of different methods. As a consequence, it can be difficult to compare results from dynamic FC studies.

Reliability and reproducibility in dynamic FC studies may mean several things. Reliability may pertain to the question: When we have several scanning sessions of a subject, do we reliably find the same patterns of dynamic FC in all sessions? We consider this the question of test-retest reliability. Studies addressing this question have shown that test-retest reliability of many dynamic FC methods is low, particularly for sliding window-based approaches (Choe et al., 2017; Zhang et al., 2018). To what extent this is due to methodological limitations or systematic within-subject variability is an important topic for future research into dynamic FC reliability (Geerligs et al., 2015).

Reliability may also refer to the question: How reliable is the method in detecting dynamic FC? This question can be addressed by simulation studies or by studying how biased a method is by secondary parameters. For instance, Ahrends et al. (2022) showed how a HMM's ability to detect temporal changes in FC depends on several parameters that are determined by preprocessing strategies. Optimising preprocessing strategies (as described in the previous section *Preparing data for dynamic FC analyses*) is therefore an important step in improving reliability.

Other studies have investigated the reproducibility of dynamic FC methods. Reproducibility may refer to the question: Do we find consistent, canonical patterns of resting-state dynamic FC across multiple datasets? For instance, Abrol et al. (2017) tested this question across datasets from different scanning sites. Using two different state-based approaches, they could show that the same basic connectivity patterns emerge in all datasets, indicating that general patterns in dynamic FC are reproducible and robust to variations in the specific datasets.

In some methods, reproducibility also entails the question: Do we get the same results when we run the model several times? In models that depend on random initialisation, solutions may to some extent be different every time the model is run (Vidaurre et al., 2019). This issue may also be referred to as robustness. Vidaurre et al. (2018) have shown that dynamic FC results from HMMs which use random initialisation for the inference are reproducible across different runs of the model to some extent, but that the similarity between runs depends on the dataset. Robustness may also refer to the issue of how the model deals with noise, e.g., in terms of noisy fMRI recordings, or in terms of outliers in the dataset, i.e., does the estimation of dynamic FC in a group of subjects suffer if single subjects display anomalous patterns of dynamic FC. This question is only starting to be addressed for different dynamic FC methods.

Assessing and improving reliability, reproducibility, and robustness of dynamic FC should be a central task in future dynamic FC research. This may not only improve the estimation of dynamic FC but also make methods and results more comparable.

## 5.2 Predicting behaviour and individual traits from dynamic FC

Time-varying FC measures have been shown to be highly individual, suggesting that they could be used for "fingerprinting" an individual (J. Liu et al., 2018). Beyond explaining behaviour, there is thus growing interest in using dynamic FC to predict individual traits and behaviour with a potential to develop dynamic FC-based biomarkers.

So far, studies using features derived from dynamic FC were able to predict various phenotypes and behavioural measures at moderate to fair accuracy, such as cognitive performance or intelligence (J. Liu et al., 2018; Sen & Parhi, 2021; Shine et al., 2019), task performance (Fong et al., 2019), or sleep quality (Zhou et al., 2020), as well as to classify clinical diagnoses, such as schizophrenia (Bhinge et al., 2019; Cetin et al., 2016) or post-traumatic stress disorder (PTSD) (Ou et al., 2015). This is an interesting avenue, because dynamic FC contains information complementary to structural measures or static FC (Ge et al., 2017; Liégeois et al., 2019; Vidaurre et al., 2021), and so it may be able to predict distinct aspects of behaviour. It has in fact been shown that dynamic FC can outperform structural and static measures in predicting behaviour (Saha et al., 2021; Vergara et al., 2020; Vidaurre et al., 2021) and in classifying neurological and psychiatric disease (Jin et al., 2017; Rashid et al., 2016). For a review on prediction and classification from static and dynamic FC in clinical applications, see Du et al. (2018).

But how can we best use the information from dynamic FC models in the context of prediction problems? One option is to select a feature of interest, such as the occurrence of a specific network, as a predictor. This is the approach taken in most studies so far. However, not only is it not obvious how to choose the relevant feature in the context of neuroimaging (Du et al., 2018; Wolfers et al., 2015), doing this would also mean losing information that was originally contained in the dynamic FC model. Instead, we may want to use all information from subject-level dynamic FC models in a linear predictive model to predict these subjects' phenotype. While static measures can relatively straightforwardly (with the exception of geometric constraints) be used in linear predictive models, the temporal dimension and complexity of dynamic FC models make this more difficult.

Developing new approaches to predicting from dynamic FC models is a central goal in current dynamic FC research. One promising avenue are kernel methods (Shawe-Taylor & Cristianini, 2004). These methods have several advantages: They allow working with high-dimensional features such as dynamic FC features in a computationally efficient way by using a subject-by-subject similarity matrix for the prediction rather than the features themselves. Kernels can be constructed in a way that preserves the structure of the features (e.g. the scales and relationships between parameters of an underlying dynamic FC model). Furthermore, kernels can be used straightforwardly in linear prediction models or classifiers. Indeed, by using kernel functions that apply a nonlinear transformation to the data, these linear prediction models or classifiers can be used to detect nonlinear decision boundaries. This elegantly combines the advantages of linear models, which are computationally efficient, easy to implement, and can be more readily interpreted, with nonlinear models, which allow finding more complex relationships and patterns in data. Kernel methods also open the door to predicting from multiple modalities, such as structural and functional, static and dynamic information, or MRI and MEG/EEG (Engemann et al., 2020; Schouten et al., 2016). Using a Multi-kernel learning (MKL) approach (Gönen & Alpaydın, 2011), separately constructed kernels for each modality can be combined within a single prediction model. This approach has been shown to improve prediction accuracy, for example when combining MRI and MEG (Vaghari et al., 2022) or several levels of FC estimation (Zhang et al., 2017).

# 6  Conclusions

In this chapter, we have given a practical introduction to different methods of dynamic FC. We also discussed data preparation and model evaluation, as well as some challenges and perspectives in the dynamic FC research field. As we have shown, there is a variety of methods to estimate dynamic FC, each with unique advantages but also shortcomings. Recent years have seen an explosion of dynamic FC studies, indicating the great potential of a dynamic view of FC in answering specific research questions and in contributing to our general understanding of brain function. However, we also stress the importance of methodological rigour in the further development of these methods. For instance, dynamic FC models may be more heavily biased by preprocessing strategies than other methods, because they are more sensitive to temporal noise. Another not fully resolved issue is the comparison to null models that assume stationarity, an important limitation that should be kept in mind in both methods development and the conceptualisation of dynamic FC. Going forward, we hope to see further work on these issues, as well as explicit testing and improvement of the reliability and reproducibility of the different methods.